# Noise-reducing attention cross fusion learning transformer for histological image classification of osteosarcoma

Liangrui Pan[1], *IEEE Student Member*, Hetian Wang[1], Lian Wang[1], Boya Ji[1], Mingting Liu[1], Mitchai Chongcheawchamnan[2], IEEE *Senior Member*, Jin Yuan*[1], Shaoliang Peng*[1]

***Abstract**—The degree of malignancy of osteosarcoma and its tendency to metastasize/spread mainly depend on the pathological grade (determined by observing the morphology of the tumor under a microscope). The purpose of this study is to use artificial intelligence to classify osteosarcoma histological images and to assess tumor survival and necrosis, which will help doctors reduce their workload, improve the accuracy of osteosarcoma cancer detection, and make a better prognosis for patients. The study proposes a typical transformer image classification framework by integrating noise reduction convolutional autoencoder and feature cross fusion learning (NRCA-FCFL) to classify osteosarcoma histological images. Noise reduction convolutional autoencoder could well denoise histological images of osteosarcoma, resulting in more pure images for osteosarcoma classification. Moreover, we introduce feature cross fusion learning, which integrates two scale image patches, to sufficiently explore their interactions by using additional classification tokens. As a result, a refined fusion feature is generated, which is fed to the residual neural network for label predictions. We conduct extensive experiments to evaluate the performance of the proposed approach. The experimental results demonstrate that our method outperforms the traditional and deep learning approaches on various evaluation metrics, with an accuracy of 99.17% to support osteosarcoma diagnosis.*

*Index Terms*—**osteosarcoma, denoise, transformer, feature fusion, framework, diagnose,**

## I. INTRODUCTION

Osteosarcoma is a malignant tumor that originates from the bone, and grows rapidly to form tumor bone-like tissue [1]; it is a common orthopedics disease. Generally, osteosarcoma easily occurs at the lower end of the femur, upper end of the tibia, and the upper end of the humerus, especially around the knee joint. In the population, osteosarcoma tends to occur in adolescents and children [2], and its symptoms include mild local bone pain, redness, and fever at the tumor site. Persistent pain by osteosarcoma affects patient movement, and thus it is one of the most important tumors that seriously affects labor productivity and even threatens life. Therefore, early diagnosis and treatment have particular significance.

[T]his work was supported by NSFC Grants 61772543, U19A2067; Science Foundation for Distinguished Young Scholars of Hunan Province (2020JJ2009); National Key R&D Program of China 2017YFB0202602, 2018YFC0910405, 2017YFC1311003, 2016YFC1302500; Science Foundation of Changsha kq2004010; JZ20195242029, JH20199142034, Z202069420652; The Funds of Peng Cheng Lab, State Key Laboratory of Chemo/Biosensing and Chemometrics; the Fundamental Research Funds for the Central Universities, and Guangdong Provincial Department of Science and Technology under grant No. 2016B090918122.

Liangrui Pan, Jin yuan, Hetian Wang, Lian Wang, Boya Ji, Mingting Liu, Shaoliang Peng are with the College of Computer Science and Electronic Engineering, Hunan University, Hunan University, Chang sha, 410000, China (e-mail: lip141772@gmail.com; {yuanjin; gorgor; lianwang; byj; Mingting; slpeng}@hnu.edu.cn).

Mitchai Chongcheawchamnan are with the Department of Electrical Engineering, Faculty of Engineering, Prince of Songkla University, Son gkhla 90110, Thailand (e-mail: mitchai.c@psu.ac.th).

Existing diagnostic techniques, including MRI, ultrasound, computer tomography (CT) and positron emission tomography (PET), have a crucial role in tumor detection [3]–[6]. However, when these techniques cannot yield an accurate judgment, doctors prefer to extract tissue samples from tumor for further analysis. Concretely, the extracted samples will be transformered into slides or smears, and then stained to show certain details of the cells, which is usually time-consuming and causes great pain patients. Therefore, the development of automatic detection technology for osteosarcoma has great value.

Recently, automatic analysis algorithms of microscopic images by computers have become the primary tool for cancer detection. These algorithms also provides a feasible solution for radiologists and pathologists to automatically detect benign and malignant tumors based on images [7]. However, compared with doctors' predictions, automatic analysis algorithms usually have lower performance in both efficiency and accuracy, which greatly limits their applications. Fortunately, the advent of the artificial intelligence era introduces hope to the diagnosis of this tumor, and clinical applications have become more realistic [8]. Machine learning algorithms have been developed to extract visual features from histological images and to calculate similarities between two different images, which promotes the continuous progress of diagnostic models [9], [10].

This paper aims to explore an automatic analysis algorithm for the diagnosis of osteosarcoma. Instead of adopting traditional image classification networks such as NB, SVM, and CNN [12]–[21], [23], [24], our research focuses on improving the structure of the vision transformer to enhance the image classification performance for osteosarcoma prediction. Vision transformers demonstrate superior performance in natural image classification and can support large-scale parallel computing. However, current transformer–based approaches directly receive natural images without any noise preprocessing, which is not suitable for histological images since noise information would degrade the performance for osteosarcoma prediction. Moreover, how to integrate multiscale image features in transformers has rarely been explored for histological image classification, which has a deep impact on classification accuracy.

Motivated by this finding, this paper proposes a novel image classification framework that is based on vision transformers by integrating a noise reduction convolutional autoencoder and feature cross fusion learning layer (NRCA-FCFL). First, the approach preprocesses images by a noise reduction convolutional autoencoder to reduce noise in histological images. Second, two independent vision transformers (ViTs) transfer the local image information of different scales to the feature cross fusion learning layer, which aims to effectively fuse both features according to the CLS tokens for performance improvement. Last, a nonlinear residual neural network is adopted as the classifier to output labels for each histological image. We conducted extensive experiments on osteosarcoma data from UT Southwestern/UT Dallas for viable and necrotic tumor assessment. The experimental results demonstrate that our method consistently performs well. The main contributions are summarized below:

1. We propose a novel image classification framework NRCA-FCFL, based on a vision transformer for tumor survival and necrosis assessment. The integration of the noise reduction convolutional autoencoder could effectively reduce noise in histological images to boost the performance
2. We introduce the feature cross fusion learning layer to effectively integrate both features from an image with different patch scales. Two CLS tokens are adopted to exchange information to better extract robust visual features.
3. We fully explore the performance of existing machine learning methods on osteosarcoma tissue image analysis tasks, and offer a baseline model for scholars to carry out follow-up research. Extensive experiments are conducted and demonstrate the superiority of the proposed approach.

## II. RELATE WORK

The diagnosis of osteosarcoma has always been a complex problem in the medical field. Many researchers use machine learning algorithms to explore the diagnosis methods of osteosarcoma [24]. Zhi et al. employed logistic regression, the support vector machine (SVM) and the random forest (RF) to classify metabolomics data of healthy controls and patients with benign tumors or osteosarcoma[25]. Logistic regression, the support vector machine and the random forest have accuracy rates of 88%, 90%, and 97%, respectively, which successfully distinguished healthy controls and tumor cases. The four pseudogene classifier identified by Feng et al. through machine learning can be applied as a new prognostic marker for the survival of osteosarcoma; its AUC value reaches 0.878 [26]. Harish et al. utilized machine learning and deep learning models to evaluate viable and necrotic tumors from the entire slide image of osteosarcoma [27]. Bingsheng et al. proposed a noninvasive and accurate method to assess the necrosis of osteosarcoma after NACT by combining mpMRI with machine learning. The results show that machine learning can more accurately distinguish tumor necrosis and tumor survival [28].

With abundant computing resources, deep learning has gradually become the preferred method of image classification [29], [30]. Rashika et al. constructed a convolutional neural network to analyze the histopathology of osteosarcoma; its diagnosis accuracy rate was 92% compared with the AlexNet, LeNet and VGGNet models [31]. DM et al. selected six popular deep learning models to explore the best osteosarcoma classification model. Among them, the VGG19 model has an accuracy rate of 96% in binary and multiclass classification tasks [32]. Yu et al. designed a deep model with a conjoined network (DS-Net). Through experiments on histological slides of osteosarcoma stained with hematoxylin and eosin (H&E), DS-Net can achieve a 95.1% average accuracy rate [33]. David et al. designed an effective labeling method for osteosarcoma treatment response evaluation based on deep learning. The CNN model uses only 7 hours of annotation training and can successfully estimate the necrosis rate within the expected interobserver variation rate of nonstandardized manual surgical pathology tasks [34].

Because of the successful application of the transformer to the image classification field, the performance of the transformer has been better than the best neural network on a large-scale data set. Xiyue et al. proposed a hybrid model (TransPath), which is pretrained in an SSL manner on massively unlabeled histopathological images to discover the inherent image property and to capture domain-specific feature embedding[35]. Hang et al. presented a novel embedded-space MIL model that is based on a deformable transformer (DT) architecture and convolutional layers, termed DT-MIL[36]. Ziyang et al. proposed a new method for survival prediction, which is named SeTranSurv. SeTranSurv extracts patch features from WSIs through self-supervised learning and adaptively aggregates them according to their spatial information and correlation between two patches using the transformer[37]. Presently, the transformer model has been widely employed in the classification task of medical images. However, the transformer model still has certain shortcomings, mainly including ① the need for a large amount of training data; ② excessively large training parameters and model; ③ inability of the transformer model based on the perceptron to surpass the model of CNN or the combination of CNN and transformer in terms of noise reduction and classification performance.

## III. MATERIALS AND METHODS

### A. Materials

From 1995 to 2015, a team of clinical scientists at the University of Texas Southwestern Medical Center in Dallas collected archive samples from 50 osteosarcoma patients at Dallas Children's Medical Center to create this dataset. This dataset consists of histological images of osteosarcoma stained with hematoxylin and eosin (H&E). According to the primary cancer type in each image, the images are labeled as nontumor, viable tumor, and necrosis by two medical experts, where each image is only labeled by one expert. As a result, the dataset contains 1,144 images with a 1,024 × 1,024 resolution and 10X magnification, including 536 (47%) nontumor images, 263 (23%) necrotic tumor images, and 345 (30%) surviving tumor images. Fig. 1 shows three examples of nontumor, necrotic tumors and viable tumors. To obtain additional samples, we split each 10X magnification image into 16 40X magnification images, each of which has a 256 × 256 resolution. As a result, the dataset expands to 18,304 images, where 750 slices are taken as the testing set, and the remaining slices are divided into the training set and the validation set at a ratio of 0.8:0.2. The raw data and their detailed descriptions can downloaded from the website[1].

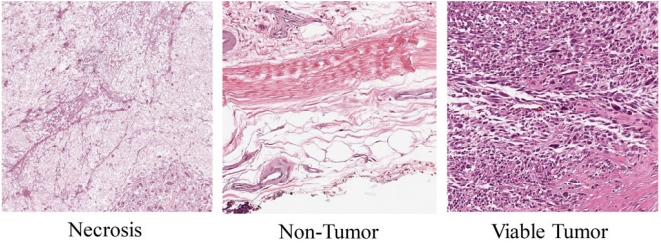

Fig. 1 An illustration of three histological images with necrosis, nontumor, and viable tumor for osteosarcoma.

### B. Proposed method

Fig. 2 demonstrates the framework of our NRCA-FCFL. Given an input image, first, a noise reduction convolutional autoencoder (NRCA) reduces its noise to generate a denoising image. Second, the denoising image is divided into two groups of image patches with different sizes, and then each group is linearly projected to a token sequence with an additional classification token (CLS). Third, two token sequences are fed to the multiscale transformer encoder to extract robust visual features, where each multiscale transformer consists of two branches to separately process each token sequence, and then fuses them by using the feature cross fusion learning layer. Last, the generated fusion feature is passed to the residual neural network for classification. Next, we will elaborate the details of the noise reduction convolutional autoencoder, multiscale transformer and residual neural network.

[1] https://wiki.cancerimagingarchive.net/pages/viewpage.action?pageId=52756935

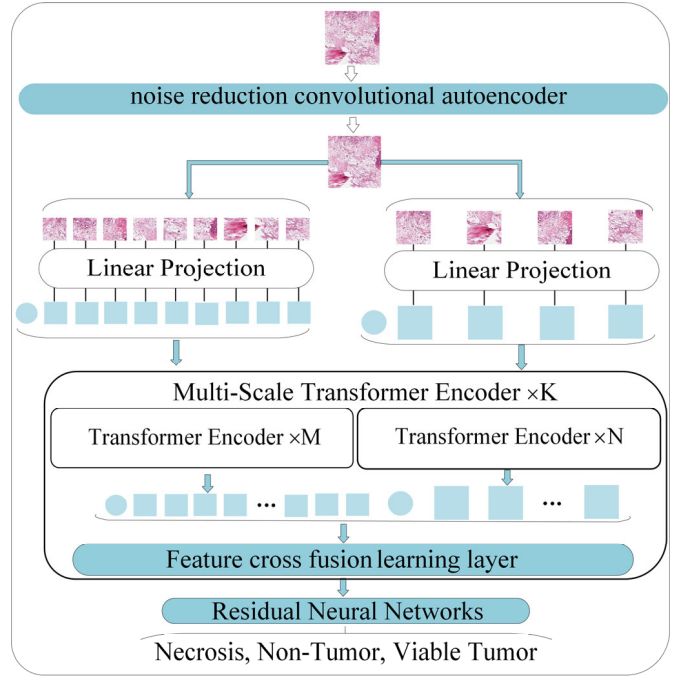

Fig. 2 An overview of the NRCA-FCFL framework.

#### 1) Noise reduction convolutional autoencoder

Different from natural images, data slicing and stacking for histological images may lead to a large amount of information loss. Moreover, a histological image usually contains useless noise information, which significantly affects the classification performance for osteosarcoma prediction. Therefore, our framework introduces a noisy reduction convolutional autoencoder (NRCA) to alleviate this problem, which has been proven to be effective for medical image classification [38]–[40]. As shown in Fig. 3, the NRCA consists of two components: encoder and decoder. The encoder is composed of multiple blocks, where each block adopts a convolution operation to extract features from a 2D image; and then performs pooling and normalization to filter useless information. As a result, amounts of noise information are discarded. On this basis, the decoder adopts transposed convolution to recover the image by increasing the feature map and amplifying the dominant feature information. Therefore, the recovered image contains more useful information with less noise than the original image, which would better support the following steps.

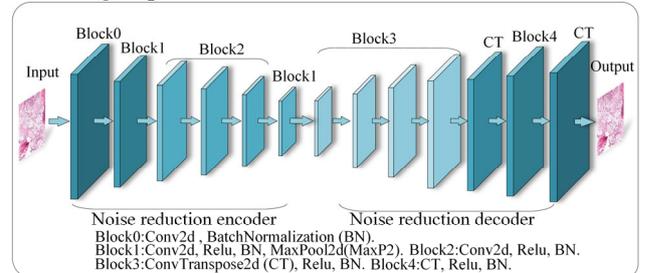

Fig. 3 An illustration of the details of the NRCA structure.

## 2) *Multiscale attention cross fusion learning transformer*

After noise reduction, images are directly transmitted to two independent ViTs for feature extraction. Since the size of the image patch will affect the classification accuracy, we propose to use two different patch sizes as the inputs of the two ViTs. After the convolution operation, image patches of different sizes undergo a linear mapping to obtain their relative position information. The difference between image processing and neural networks lies in the additional classification token (CLS), which is the primary indicator for classification [38][39]. ViT uses the CLS token that interacts with the patch mark on each ViT encoder as the final embedding. Based on this finding, we introduce the dual-channel ViT to encode small and large tokens from two independent branches. For each marker of the two branches, we add a learnable position embedding in front of the marker to learn the position information in the ViT. We then propose a feature cross fusion learning layer to fuse the image information transmitted by both ViTs. As shown in Fig. 4, our ViTs are mainly composed of K multiscale feature fusion encoders, where each encoder is primarily composed of two independent branches: (1) a small patch branch which employs many small patches with wide embedding dimensions, and (2) a large patch branch which works on large patches with small embedding dimensions. Two independent ViTs are encoded by the transformer by N and M times, and then the two branches are merged by K times; CLS tokens of the last two branches are utilized for prediction.

The most important module in our framework is the feature cross fusion learning layer, which can integrate the CLS token of one branch and the patch token of the other branch. To more effectively integrate the feature information from two independent ViTs, we use the CLS token on one branch as a proxy exchanging information between two patch tokens from the other branch, and then project it back to our branch. Since the CLS token has learned the abstract information among all patch tokens in its branch, the interaction with the patch token on the other branch helps contain information of different scales [43]. After merging, the CLS token interacts with its patch token again at the next transformer encoder, and can transfer the learned information from the other branch to its patch token, resulting in more rich information. The CLS token divides the features into three parts, and conducts self-learning and improvement under the encouragement of self-attention.

We select the feature cross fusion module with a small patch branch for analysis. First, the independent small branch collects the patch token and connects the CLS token to the corresponding patch token. Assuming that the input small patch token is $X_{patch}^l$ and that its CLS token is $X_{CLS}^h$, we express it as

$$X_{CLS}^{'h} = [f^h(X_{CLS}^h) \| X_{patch}^l], \quad (1)$$

where $f^h(\cdot)$ is the projection function for dimension alignment. The feature cross fusion mainly consists of the information exchange between modules $X_{CLS}^h$ and $X_{CLS}^{'h}$. Given the token information of the small branch, the ViT image is fused into the CLS token, and then the process of feature cross fusion is

$$y_{patch}^h = soft\max(X_{CLS}^{'h}W_x \cdot (X_{CLS}^{'h}W_y)^T / \sqrt{C/h})X_{CLS}^{'h}W_z, \quad (2)$$

where $W_x$, and $W_y$ are learnable parameters, and C and h are the embedding dimension and number of heads, respectively. Since we only use CLS in the query, the computational and memory complexity of generating the attention map in cross-attention is linear, rendering the whole process more efficient. The output after the final feature cross fusion is defined as

$$y_{CLS}^{'h} = g^h(X_{CLS}^{'h} + y_{patch}^h) + X_{patch}^h, \quad (3)$$

where $g^h(\cdot)$ is the back-projection function for dimension alignment. After feature cross fusion, features will become more complex, and the information on the CLS token carries multidimensional information that needs to be learned. We recommend adopting the self-attention mechanism to learn the features after cross fusion. The module is shown in Fig. 4, and the final output can be expressed as

$$Output_{self-Attention} = W_{V_{S,L}} Soft\max(W_{Q_{S,L}} W_{K_{S,L}} / \sqrt{d}). \quad (4)$$

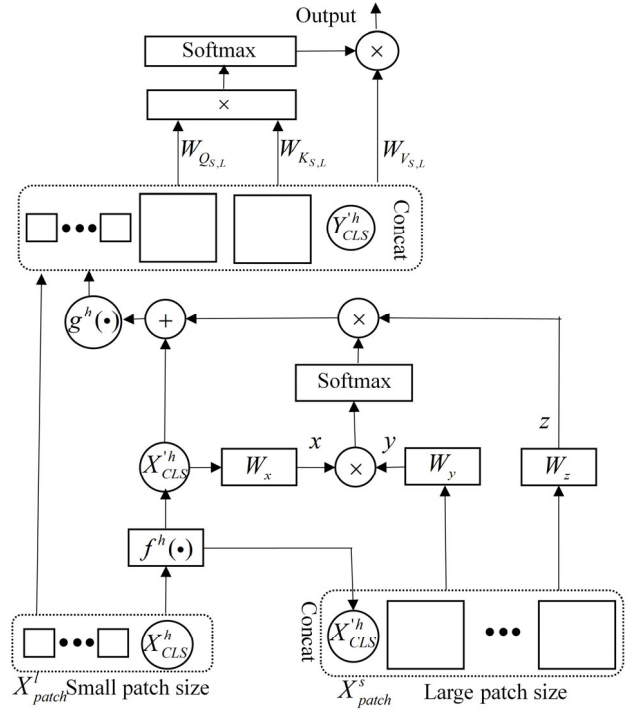

Fig. 4 The working principle of the feature cross fusion learning module. The CLS token of the small patch branch is used as the query token, which interacts with the patch token of the large patch branch. Conversely, the CLS token of the large patch branch is used as the query token, which can interact with the patch token of the small patch branch.

The self-attention mechanism is a variant of the attention mechanism that reduces the dependence on external information, and is better at capturing the internal correlations of data or features [44]. The self-attention mechanism is the complement of convolution [45], and the CLS tokens and image features after feature cross fusion are irregularly concentrated after the attention layer. Therefore, the self-attention mechanism can further effectively extract image feature information.

### 3) Residual neural network

A large number of experiments show that the nonlinear CNN classifier is better than the traditional linear MLP [46]. For the classification task of histological images, CLS tokens are the target of the classification task, and their complexity is relatively high. Therefore, we recommend using a residual neural network for classification since the residual structure could effectively prevent the overfitting of the model. Fig. 5 demonstrates the structure of the residual neural network, where softmax is adopted to output the classification label for an input image.

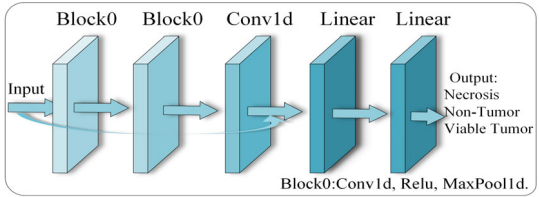

Fig. 5 The structure of the residual neural network.

### 4) Training

The cross entropy loss function is employed in our framework to update the parameters, which is expressed as follows:

$$L_{cost}(q_s, q_d) = -\sum_{i=1}^{N} q_d^{(i)} \log q_s^{(i)}, \quad (5)$$

where $q_s$ and $q_d$ represent predicted label probability and the true label probability, respectively. In our framework, all the components are jointly trained by using the loss function, and thus errors could be propagated through the whole pipeline to optimize each component.

## IV. EXPERIMENTS

### A. Experimental steps

Considering the insufficiency of histological images and the label imbalance, first, we use Gaussian noise and shot noise to expand the histological images, yielding the equality of each type of sample. Second, we use data augmentation technology to expand the training set. For example, the images are flipped by 90 degrees. The translation distance in width and height is the length or width of the image multiplied by 0.1. The cutting angle in the radian counterclockwise direction is 0.2. Points outside the input boundary are filled with abcd according to the given pattern. Third, NRCA-FCFL is trained on the training set to generate a diagnostic model. We employ RF, NB, KNN, SVM, VGG19, Xception, MobileNetV2, Resnet50, DenseNet201, InceptionV3, EfficientNetB7 and InceptionResNetV2 to train multiple diagnostic models. We then evaluate the performance of these models on the testing set.

### B. Implementation Details

**NRCA-FCFL:** We implemented our model in PyTorch and trained the model on 2 GPUs (NVIDIA V100) with a batch size of 64. The patch sizes in NRCA-FCFL are 12 and 16, respectively, and the corresponding embedding dimensions are 192 and 384. We trained the model for 500 epochs in total, and set the initial learning rate to be 0.0001 and decay the learning rate by a factor of 0.5 in every 30 epochs. AdamW is used as the optimizer.

**Other models:** For RF, the number of trees is set to 10, the maximum depth is 13, and the out-of-bag samples are utilized to estimate the generalization score. The floating-point number settings of NB is smooth and the algorithm has no prior probability. For KNN, we set $k = 6$ as the number of clusters. The SVM algorithm employs a Gaussian kernel function. NB, RF, KNN, SVM are required to reduce the image dataset from a three-dimensional array to a one-dimensional array, and then train the models. The image size is 224*224 for the deep learning approaches, and the dataset is randomly shuffled. We trained all the models for 500 epochs in total, and set the initial learning rate to be 0.0001 and decayed the learning rate by a factor of 0.5 in every 30 epochs. Adam is the optimizer of the deep learning models. All models were trained with an NVIDIA V100.

### C. Evaluation

We employ a variety of evaluation metric to evaluate the performance, including precision, recall, F1, accuracy, and AUC. Precision represents the proportion of truly positive samples among all identified ones by the model, while recall represents the ratio of positive samples correctly identified by the model to the total ones. In general, the higher of recall indicates the more positive samples to be predicted correctly by the model. F1 score is also called BalancedScore, which is understood as the weighted average of precision and recall. The best value of F1 score is 1 and the worst value is 0. AUC (Area Under Curve) is defined as the area under the ROC curve, and its value is less than or equal to 1. It is widely used as the evaluation criterion in many cases, and the larger AUC means the better performance.

In our experiment, all the experimental results are calculated by using cross validation, where we randomly divide the training set, validation set, and testing set 10 times to train and test a model. The performance is estimated by averaging the their results.

### D. Experimental Results

#### 1) Comparison with state-of-the-art models

Table I demonstrates the performance among different approaches measured by precision, recall, F1 score, accuracy and AUC values. Concretely, the performance of SVM reaches 0.9352, 0.9307, 0.9295, and 0.9307 on precision, recall, F1 score, and accuracy, respectively, which is better than the other traditional algorithms including NB, RF and KNN. The superiority of the SVM stems from the use of kernel function which could map low features to a new feature space, which is much high for better classification.

InceptionResNetV2 model achieves the best performance among the eight deep learning models, with 0.9726 on precision, 0.972 on recall, F1 scores and accuracy, and 0.99 on AUC. However, this model has a large model size of 623 MB, which will significantly increase the training and testing costs. Moreover, it is demonstrated that the deep learning models usually have better performance than the traditional approaches. This is because the deep learning approaches could extract robust visual features to conduct a joint training in feature extraction and prediction, while the traditional methods separately extract hand-drafted features and then make a label prediction.

TABLE I PERFORMANCE COMPARISON AMONG DIFFERERNT APPROACHES MEASURED BY PRECISION, RECALL, F1 SCORE, ACCURACY AND AUC VALUES.

|  | Models | Precision | Recall | F1 score | Accuracy | AUC | Size(MB) |
|---|---|---|---|---|---|---|---|
| **Traditional algorithm** | NB | 0.6102 | 0.585 | 0.5604 | 0.5933 | 0.64 | 7 |
|  | RF | 0.7791 | 0.7539 | 0.7471 | 0.76 | 0.91 | 1 |
|  | KNN | 0.8868 | 0.8758 | 0.8767 | 0.8783 | 0.77 | 2690 |
|  | SVM | 0.9352 | 0.9307 | 0.9295 | 0.9307 | 0.98 | 2080 |
| **Deep learning algorithm** | EfficientNetB7 | 0.7001 | 0.6187 | 0.5889 | 0.6187 | 0.43 | 734 |
|  | VGG19 | 0.8484 | 0.7973 | 0.8001 | 0.7973 | 0.95 | 230 |
|  | Xception | 0.9011 | 0.8987 | 0.8987 | 0.8987 | 0.98 | 239 |
|  | InceptionV3 | 0.7811 | 0.7333 | 0.7251 | 0.7333 | 0.9 | 250 |
|  | MobileNetV2 | 0.6527 | 0.5933 | 0.5214 | 0.5933 | 0.96 | 26 |
|  | DenseNet201 | 0.6924 | 0.608 | 0.5766 | 0.608 | 0.79 | 210 |
|  | ResNet50 | 0.9085 | 0.8987 | 0.9002 | 0.8987 | 0.98 | 270 |
|  | InceptionResNetV2 | 0.9726 | 0.972 | 0.972 | 0.972 | 0.99 | 623 |
|  | **NRCA-FCFL** | 0.9934 | 0.9893 | 0.9903 | 0.9917 | 0.99 | 130 |

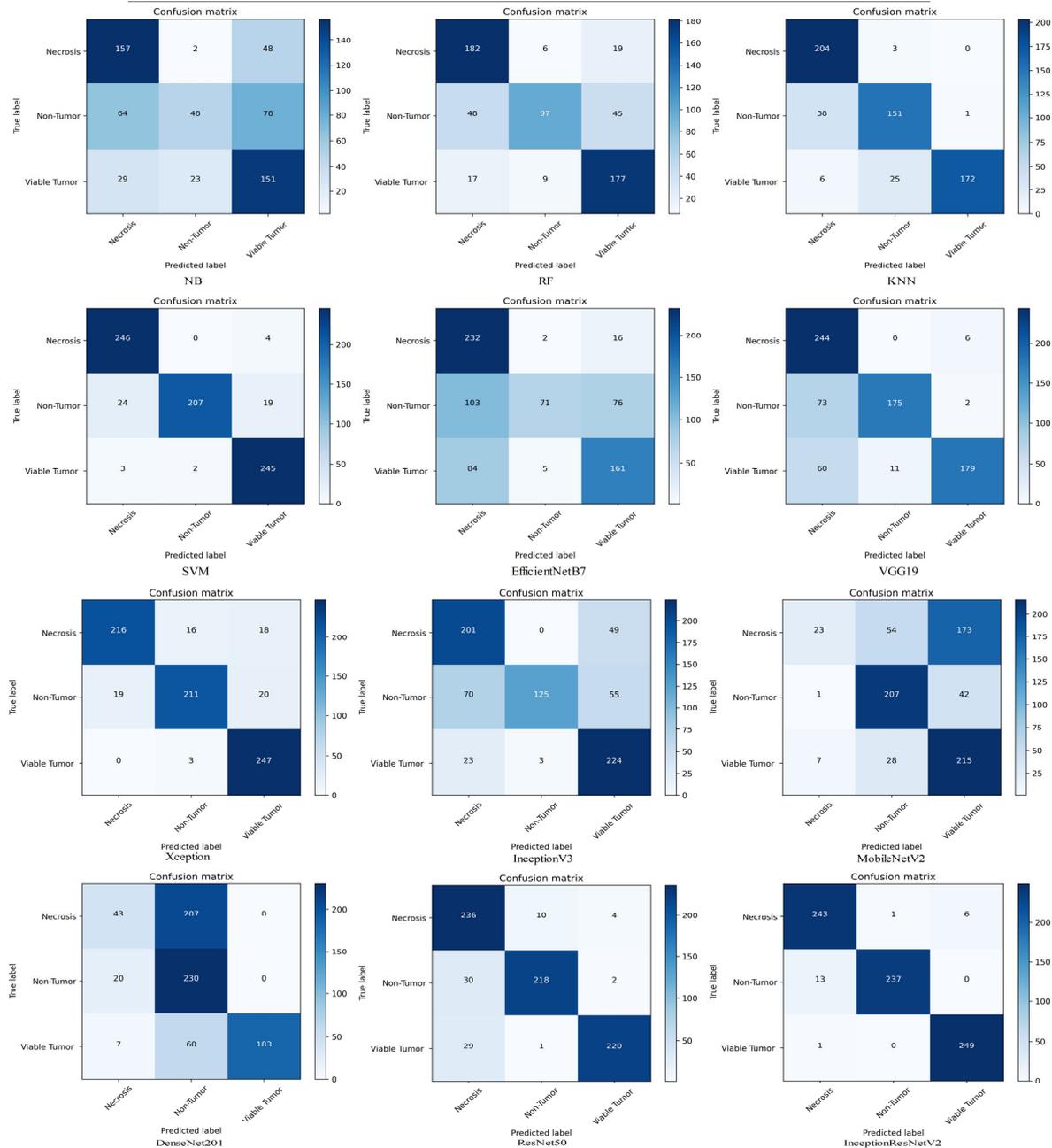

Fig. 6 Illustration of different confusion matrices by different approaches on testing dataset.

Compared to the previous methods, our NRCA-FCFL achieves the best performance with 0.9894, 0.9893, 0.9893, 0.9893 on precision, recall, F1 score, and accuracy, respectively, and the AUC value rises to 0.99, which demonstrates the superiority of our approach in image classification. This significant improvement comes from the use of the self-attention mechanism for visual feature extraction as well as the proposed noise reduction component and multi-scale feature fusion learning layer. Moreover, we discover that this high performance does not require a large model size, with only 130 MB, which is much less than most deep learning models. This is because the used transformer is a weight sharing structure, and thus the repetitive multiple transformers would not increase the model size.

Fig. 6 and 7 adopt the confusion matrix to illustrate the detailed classification results on each category. Each column represents the sample predictions on each category by the model, and each row indicates the true labels of samples on each category. It is demonstrated that NB, EfficientNetB7, MobileNetV2, and DenseNet201 have serious misjudgments in a certain category, and KNN, SVM, Xception, ResNet50, and InceptionResNetV2 have better predictions. Comparatively, NRCA-FCFL achieves the least misclassification with only six nontumor samples misclassified as necrosis and one necrosis/viable tumor misclassified as nontumor. This is because our approach could filter useless noise information in histological images and extract robust visual features.

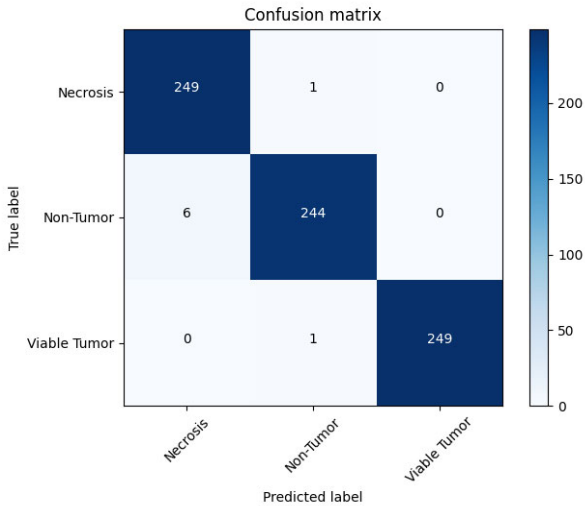

Fig. 7 Illustration of the confusion matrix by NRCA-FCFL on testing dataset.

Fig. 8 and 9 adopt the ROC curve to illustrate the classification performance of different approaches, where the steeper of the curve indicates the better performance of the approach. For each approach, we show five ROC curves, including micro-average, macro-average, necrosis, nontumor, and viable tumor. It is demonstrated that the ROC curve of nontumor by is below the diagonal, which indicates that the NB model is not suitable for detecting osteosarcoma. All the ROC curves by the SVM model are close to the upper left corner, indicating that the accuracy of diagnosis is quite high. For the deep learning model, the ROC curve of the EfficientNetB7 model is mostly parallel to the diagonal, indicating that the model has a poor diagnostic effect. The ROC curves of Xception, ResNet50, and InceptionResNetV2 models are close to the upper left corner. These models certainly have good diagnostic performance. The other models, such as VGG19, InceptionV3, MobileNetV2, and DenseNet201, have partial ROC curves close to the upper left corner and may cause misjudgments for diagnosis. For our approach, all the ROC curves are close to the upper left corner, indicating the high accuracy for diagnosis. This high performance stems from the strong ability of vision transformer with noise reduction and feature fusion components.

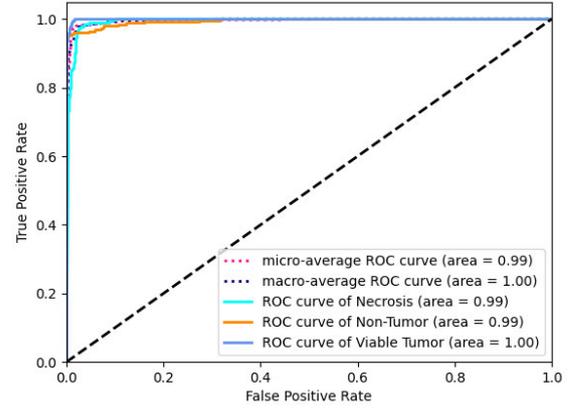

Fig. 9 Illustration of the ROC curve by NRCA-FCFL on testing dataset.

2) *Ablation studies*

**Impact of patch size.** To evaluate the influence of the patch size in NRCA-FCFL, we tried different patch sizes of $12 \times 12$, $12 \times 16$, $16 \times 16$ for two independent ViTs in our approach, and TABLE II demonstrates the results. When the patch size is $12 \times 12$, the accuracy of NRCA-FCFL arrives 98.93%, while that decreases to 98.02% and 97.56% for $12 \times 16$, and $16 \times 16$ patch sizes. This results indicate that smaller patch size would result in better performance. This is because the smaller patch size will lead to the more patch samples for feature extraction. As a result, the multi-scale transformer encoder could better extract visual features. Moreover, the smaller patch usually could provide more delicate features, which is beneficial to performance improvement.

TABLE II THE INFLUENCE OF THE PATCH SIZE BY THE NRCA-FCFL FRAMEWORK EVALUATED BY ACCURACY.

|  | Patch size | | |
| --- | --- | --- | --- |
|  | (12,12) | (12,16) | (16,16) |
| Accuracy (%) | 98.93 | 98.02 | 97.56 |

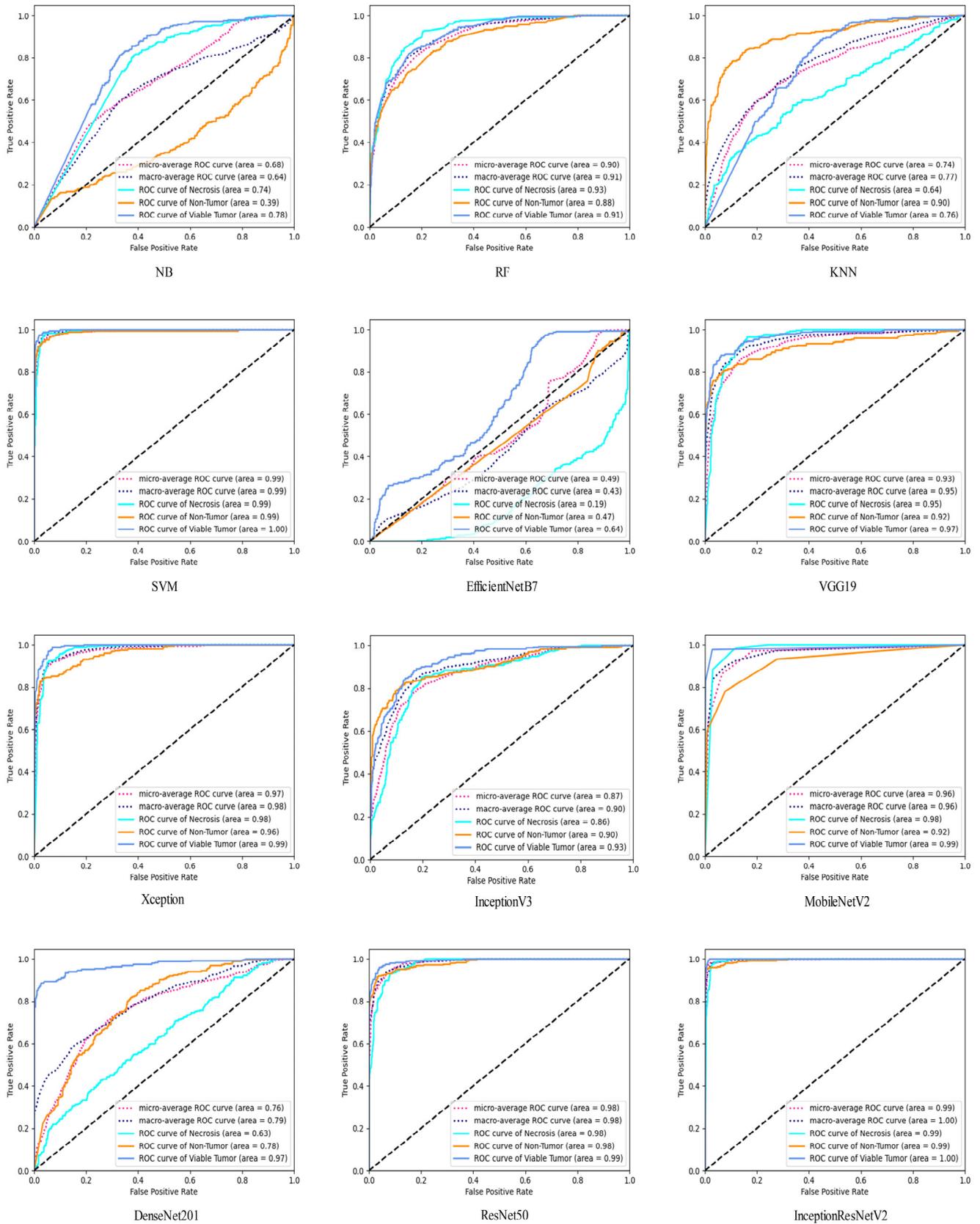

Fig. 8 Illustration of the ROC curves by different approaches on the testing dataset.

**Impact of NRCA modules.** To evaluate the effectiveness of NRCA, we conducted a set of controlled trials to explore the effect of NRCA in our approach. We trained the model separately by using NRCA and not using NRCA to calculate the accuracy. TABLE III shows the results of the two testing results. The accuracy of the NRCA-FCFL framework without using NRCA is only 97.96%, which is lower than that using NRCA with 0.97%. This result indicates that NRCA can help improve the reliability of diagnosis due to the preprocessing of images for noise reduction. Concretely, the noise of histological images is significantly reduced after convolution, pooling and normalization operations in the encoder. After that, the transposed convolution operation in the decoder restores the image to its original scale. As a result, the recovered histological images contain less noise.

TABLE III THE INFLUENCE OF NRCA IN THE NRCA-FCFL FRAMEWORK TESTED BY ACCURACY.

|  | NRCA | |
|---|---|---|
|  | Yes | No |
| Accuracy (%) | 98.93 | 97.96 |

**Impact of residual network.** In our approach, we employ a residual neural network to replace multilayer perceptron, which is widely used to perform classification in vision transformers. To evaluate the effectiveness of the residual network in our approach, we conducted a set of experiments to explore the impact of residual neural network. We trained two models by using multilayer perceptron and residual neural networks, respectively, and TABLE IV shows their testing results. The experiment found that the accuracy of the NRCA-FCFL framework by using multilayer perceptron is only 98.46%, which is lower than that of the NRCA-FCFL framework with the residual neural network by 0.47%. This is because the residual network could skip connection and set up a bridge for exchanging spatial information at different levels of images with different resolutions. It also provides a shortcut for back-propagating gradient shortcut descent for efficient training and classification.

TABLE IV THE INFLUENCE OF RESIDUAL NETWORK IN THE NRCA-FCFL FRAMEWORK TESTED BY ACCURACY.

|  | Residual neural network | |
|---|---|---|
|  | Yes | No |
| Accuracy (%) | 98.93 | 98.46 |

## V. APPLICATIONS AND LIMITATIONS

Benefitting from the high accuracy of 99.17% by NRCA-FCFL in slice prediction, we can use to the trained model to accurately classify osteosarcoma histological slices, including predicting categories of all the small sections in a whole tissue section, counting the number of different types of sections, and calculating tumor-stromal ratio (tumor section: stromal section) and tumor cell ratio (tumor section: All slices). Therefore, this approach provides a convenient tool for doctors to analyze the microenvironment of tumor survival, as well as reduce the pains of patients in biological testing.

There are two limitations of our work. First, the number of publicly available osteosarcoma samples is small, and thus we sliced the original samples to construct the training set, validation set and testing set. Therefore, we have not sufficient samples to train the model and test the performance. The achieved high performance is only tested on a limited number of testing samples, and maybe degraded in practice. Second, NRCA-FCFL has a large model size, and requires much training time with a low detection speed on lightweight devices. Therefore, the hardware requirements for model's running are relatively high, which brings increasing costs in real cases.

## VI. CONCLUSION

This study proposed an image classification framework named NRCA-FCFL to classify histological images for osteosarcoma prediction. Our approach employs vision transformer as our backbone and focuses on filtering noisy information in histological images and extracting robust visual features for performance improvement. Technically, we design a novel noise reduction convolutional autoencoder for noise filtering, and feature cross fusion learning layer to effectively integrate visual features from two types of image patches. The architecture is an end-to-end structure, and all the components could be jointly trained to boost the performance. We conducted extensive experiments on a public dataset. The experimental results demonstrate that all the introduced components are effective, and our model consistently outperforms the state-of-the-art approaches.